\documentclass{article}
\usepackage{arxiv}
\usepackage[utf8]{inputenc}
\usepackage[T1]{fontenc}
\usepackage[square,numbers,sort&compress]{natbib}
\usepackage{hyperref}
\usepackage{url}
\usepackage{booktabs}
\usepackage{amsfonts}
\usepackage{microtype}
\usepackage{graphicx}
\usepackage{threeparttable}
\usepackage{multirow}
\usepackage{multicol}
\usepackage{xcolor}
\usepackage{subcaption}
\usepackage[font=small,labelfont=bf]{caption}
\usepackage{amsmath,amsfonts,bm}
\usepackage{amssymb,amsthm}
\usepackage{enumitem}
\usepackage{adjustbox}
\usepackage{float}
\usepackage{array}
\usepackage{makecell}
\usepackage[perpage]{footmisc}
\usepackage{wrapfig}
\usepackage{tcolorbox}
\usepackage{tikz}
\tcbuselibrary{breakable, skins}
\usetikzlibrary{arrows.meta,positioning,fit,calc}
\usepackage{cleveref}
\usepackage{titlesec}
\usepackage{makecell}

\setlist[itemize,1]{leftmargin=\dimexpr 18pt}
\setlist[enumerate,1]{leftmargin=\dimexpr 18pt}

\titlespacing*{\section}{0pt}{1.2\baselineskip}{0.7\baselineskip}
\titlespacing*{\subsection}{0pt}{0.9\baselineskip}{0.45\baselineskip}
\titlespacing*{\subsubsection}{0pt}{0.7\baselineskip}{0.35\baselineskip}

\newcommand{\steprealtime}{\mbox{StepAudio} 2.5 Realtime}
\newcommand{\mtpmodel}{MTP-5}

\title{
\raisebox{-0.18\height}{\includegraphics[width=0.045\textwidth]{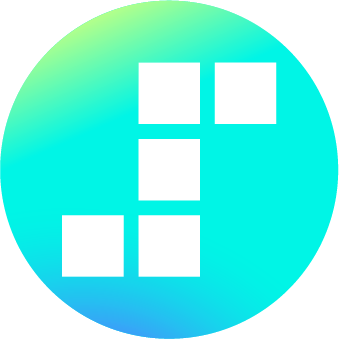}} 
StepAudio 2.5 Technical Report
}

\author{\vspace{1em} StepFun-Audio Team}

\renewcommand{\headeright}{StepFun-Audio Team}
\renewcommand{\undertitle}{}
\renewcommand{\shorttitle}{}

\begin{document}
\large

\maketitle

\begin{abstract}

Unified audio-language modeling has emerged as a prominent trend in modern speech systems, promising to bring the reasoning capabilities of large language models to auditory tasks. However, existing unified foundations often struggle to match the depth of specialized systems across automatic speech recognition (ASR), text-to-speech synthesis (TTS), and realtime spoken interaction. Bridging this gap remains an open challenge. This report presents StepAudio 2.5, a unified audio-language foundation model that matches or exceeds specialized systems across all three capabilities. \textbf{Rather than treating these tasks as architecturally distinct, we operate on the premise that once text and audio share a multimodal representational space, task specialization becomes a matter of operational regimes: data construction, optimization targets, and decoding constraints.} Guided by this insight, we advance the post-training paradigm from standard supervised learning to task-tailored Reinforcement Learning from Human Feedback (RLHF), using it as the primary mechanism to define complex optimization targets. We leverage this RLHF-centric alignment, alongside specialized decoding, to shape a shared backbone into three distinct operational modes. Concretely, the ASR branch advances transcription efficiency via verifiable multi-token decoding; the TTS branch achieves controllable, expressive synthesis through preference-based RLHF and context-rich supervision; and the Realtime branch realizes low-latency, persona-consistent dialogue via generative reward modeling within an RLHF framework. On standard benchmarks, StepAudio 2.5 achieves state-of-the-art results across ASR, TTS, and Realtime, demonstrating that a singular audio-language foundation can successfully internalize the distinct deployment objectives of speech understanding, generation, and live interaction.

\end{abstract}
\section{Introduction}
\label{sec:introduction}

Automatic speech systems are entering a period of architectural convergence, driven by the increasing dominance of large language models (LLMs): as LLMs became the standard interface for text-based reasoning, treating speech as another sequence type within the same modeling framework became a natural design choice. In automatic speech recognition (ASR), the dominant paradigm has evolved from alignment-based and encoder-decoder transduction approaches~\citep{graves2012connectionist,graves2012sequence,chan2015listen} through large-scale weakly supervised acoustic models such as Whisper~\citep{radford2023robust}, and more recently toward systems that couple strong acoustic encoders with LLM decoders~\citep{peng2026vibevoiceasrtechnicalreport, an2025funasrtechnicalreport, bai2024seed, shi2026qwen3asrtechnicalreport}. In parallel, text-to-speech (TTS) synthesis has shifted from hand-engineered pipelines toward generative modeling over increasingly abstract speech representations, with commercial systems such as ElevenLabs-v3, Minimax Speech-2.8-hd, and Gemini-Flash-TTS advancing the controllability and expressivity of synthesized speech. A third frontier has emerged in realtime conversational speech agents, exemplified by GPT-realtime, Gemini Live, and Doubao Realtime, that must understand paralinguistic signals, respond with low latency, preserve persona, and remain emotionally appropriate within an unfolding interaction. These three trajectories now meet at a common point: speech is no longer treated as a modality requiring a fully separate stack, but as another sequence type that can be mapped into and out of a shared language-centric latent space, as demonstrated by recent unified audio-language foundations~\citep{wu2025stepaudio2technicalreport, xu2025qwen3omnitechnicalreport,liu2026omni}.

The appeal of this convergence extends beyond consolidating previously separate models into a unified architecture. Traditional cascaded pipelines connect ASR, an intermediate language model, and TTS as isolated stages, inevitably discarding information when speech is reduced to a textual intermediate representation~\citep{tang2024salmonn, borsos2023audiolm, cui2025recent}. A unified audio-language foundation instead preserves speech information end-to-end, allowing paralinguistic cues, emotional state, and conversational context to directly influence recognition, synthesis, and dialogue generation~\citep{kim2024paralinguistics, wang2025freezeomni,li2026depflow,xuan2023new,deng2025multi}. Such models also directly leverage the semantic, conversational, and reasoning capabilities already developed in LLMs. Under this formulation, audio-language modeling is not merely a matter of replacing task-specific systems with a shared backbone, but of establishing a common representational substrate where information previously lost between stages remains available throughout the interaction process.

Open-source efforts such as Step-Audio 2 \citep{wu2025stepaudio2technicalreport} and Qwen3-Omni \citep{xu2025qwen3omnitechnicalreport}, alongside large-scale commercial systems including GPT-4o, Gemini~\citep{team2023gemini}, and Doubao, have all moved toward end-to-end audio-language foundations spanning ASR, TTS, and realtime spoken interaction~\citep{yan2025step,wu2025mind,wu2025chronological,zhang2026sla,cui2025recent,zhang2025mamba,liu2026code}. Despite this shared direction, simultaneously meeting the deployment requirements of all three capabilities within a single model remains challenging. ASR prioritizes accurate and efficient long-form transcription, and TTS emphasizes controllable and expressive synthesis, while realtime interaction further requires low-latency turn-taking together with persona consistency and paralinguistic responsiveness. These objectives are not naturally aligned, and existing unified systems often achieve strong performance on some capabilities while remaining behind specialized systems on others. Closing this gap remains an active focus of audio-language research.

This report presents StepAudio 2.5, building on the Step-Audio line of work \citep{wu2025stepaudio2technicalreport, tian2025step, zhang2026step} to narrow the gap between unified and specialized speech systems. The system is most naturally understood not as a collection of parallel endpoints loosely assembled around a shared name, but as a singular audio-language foundation model guided by a central thesis:

\begin{tcolorbox}[colback=gray!10, colframe=gray!10, boxrule=0pt, arc=2mm, halign=center]
\textit{Once text and audio share a well-shaped representational space, the differences among downstream tasks migrate away from architecture toward operational regimes: data, objectives, and decoding constraints.}
\end{tcolorbox}

Following this insight, we view post-training as the primary lever for shaping each capability to its specific deployment objective. Rather than treating ASR, TTS, and realtime interaction as separate engineering tracks, we refine the shared multimodal prior through a unified alignment paradigm. Crucially, we move beyond basic supervised fine-tuning (SFT) by establishing Reinforcement Learning from Human Feedback (RLHF) as the central mechanism for capturing nuanced human preferences and paralinguistic behaviors. We complement this RLHF-driven alignment with capability-specific SFT and specialized decoding strategies. Concretely, the ASR branch advances the quality-efficiency frontier by coupling the shared decoder with a verifiable multi-token decoding head, exploiting acoustic determinism to emit multiple tokens per step. The TTS branch adapts the backbone for controllable generation via semantic-to-audio alignment, integrating context-rich supervision with human-preference-driven RLHF. Finally, the Realtime branch extends the foundation toward low-latency spoken dialogue through progressive SFT for persona and paralinguistic sensitivity, followed by RLHF driven by a generative reward model and explicit interaction rubrics. On standard benchmarks across the three capabilities, StepAudio 2.5 achieves state-of-the-art results, outperforming both leading unified audio-language models and specialized systems built for individual tasks.
\section{Unified Foundation Architecture}
\label{sec:architecture}

\subsection{Shared Backbone}

The architecture follows a familiar audio-encoder--adapter--LLM-decoder pattern that has become central to audio-language modeling~\citep{wu2025stepaudio2technicalreport,xu2025qwen3omnitechnicalreport}. A frozen audio encoder converts waveform-derived features into compact acoustic embeddings. A lightweight adaptor maps those embeddings into the hidden space of a large decoder initialized from a text LLM. The decoder then operates over a unified sequence space in which conventional text tokens and newly introduced audio tokens can both appear.

This design is intentionally asymmetric. The encoder is responsible for stable acoustic abstraction, while the decoder carries the burden of semantics, context management, instruction following, and generation. Such asymmetry is not a limitation; it is the systems decision that makes the model family coherent. Once semantics live primarily in the decoder, downstream tasks can share most of the model even when their outputs differ.

\begin{figure}[h]
\centering
\includegraphics[width=1.0\linewidth]{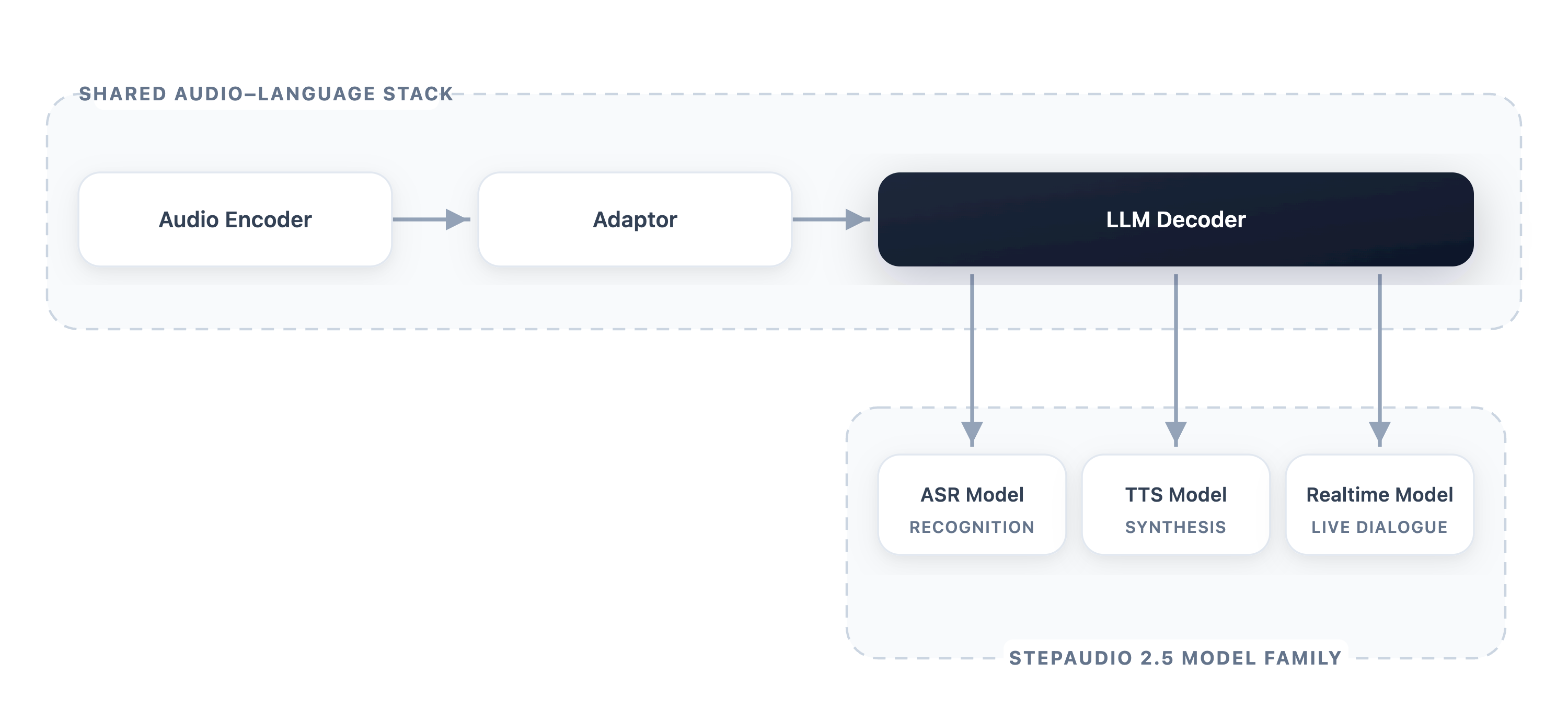}
\caption{A unified view of the StepAudio 2.5 model family. The shared audio-language stack provides the common architectural basis used to organize ASR, TTS, and Realtime, while the three systems serve different deployment goals.}
\label{fig:unified_architecture}
\end{figure}

Figure~\ref{fig:unified_architecture} summarizes the structural organization used throughout this report. At the center is the shared StepAudio 2.5 foundation model, which supports three model-level specializations: ASR, TTS, and Realtime. These three systems share the same audio-language stack while serving different deployment regimes, making the figure a compact summary of how one foundation is specialized for recognition, synthesis, and live spoken interaction.


\subsection{Task Specialization as Directional Inference}

StepAudio 2.5 supports three primary inference directions.
\begin{itemize}
\item In ASR, audio embeddings condition the decoder to generate transcript tokens. The output space is narrow, discrete, and strongly anchored by the speech signal.
\item In TTS, text and control instructions condition the decoder to generate audio tokens or intermediate audio representations. The output space is much richer, and the central challenge is not lexical correctness but faithful, natural, and expressive realization.
\item In Realtime, the model couples audio understanding and response generation under strict turn-level latency constraints, while maintaining conversational state, persona consistency, and contextual appropriateness.
\end{itemize}

This directional perspective provides a useful insight: the foundation model itself does not need separate notions of ``understanding'' and ``generation.'' It needs a single high-quality multimodal prior plus a mechanism to route supervision through different output spaces and deployment regimes. Recognition, synthesis, and realtime dialogue then become three ways of querying the same multimodal memory.

\section{Shared Data Engine and Foundation Pretraining}
\label{sec:data}

\subsection{A Common Data Production Pipeline}

StepAudio 2.5 adopts an automated data production pipeline that jointly supports speech understanding, TTS, and speech dialogue tasks. Raw audio is first processed with sound event detection (SED) and voice activity detection (VAD) to filter low-quality non-speech segments. Adjacent VAD segments are then merged and re-segmented into base samples with relatively complete semantics and suitable duration. For each audio clip within the base samples, audio-level annotations are performed, including audio quality scoring, synthetic voice detection, and speaker count labeling. At the text annotation level, dual ASR models are employed for transcription and language identification. The resulting transcripts are cross-validated with metrics such as WER, edit distance, and speech rate. Based on the ASR transcription, semantic completeness assessment and content classification are further carried out for each base sample. Finally, according to metadata, the data is categorized and graded by language, duration, semantic quality score, and audio quality score, enabling the pretraining phase to sample different data qualities for different training stages.

\subsection{Progressive Foundation Training}

StepAudio 2.5 is initialized from a textual MoE LLM and then continually pre-trained on 2.2T tokens of text and audio data. The training curriculum follows a concrete staged recipe rather than a loosely defined scaling process.

The first stage follows Step-Audio 2 and uses 3B tokens of ASR data to align speech and text feature spaces within the adaptor. During this alignment phase, both the audio encoder and the LLM remain frozen, and only the adaptor is trained. This stage establishes the initial interface through which acoustic features can be consumed by the text-native decoder.

After alignment, the model vocabulary is expanded with speech tokens, and unified multimodal training begins with a sequence length of 16K. This main pretraining mixture contains 800B tokens of text data and 800B tokens of speech data. The speech portion includes ASR, TTS, speech-to-text translation, utterance-level text-speech interleaved continuation, and speech-to-speech conversation data. In other words, the model is not exposed to audio only as transcription input, but as a general sequence modality appearing in multiple input-output configurations.

This multimodal phase is itself divided into two stages. The first is a 128B-token warmup stage designed to stabilize the newly introduced speech vocabulary and help the MoE experts adapt quickly to audio-modality data. In this stage, the adaptor, embedding layer, and output layer use larger learning rates than the base model, while the MoE router uses a smaller learning rate to reduce disruption to the text modality. The second is the main training stage, where these layer-specific learning rates are brought back in line with the base learning rate, and the MoE auxiliary loss coefficient together with the router learning rate are progressively annealed to maintain a better balance between expert utilization and top-$k$ routing probabilities.

Finally, the model enters a cooldown phase on 600B tokens of high-quality text and audio data, with the sequence length increased to 32K. In addition to the data types already used in the main training stage, this phase also introduces Audio Caption and Instruct TTS data. Relative to the earlier scaling stage, the cooldown phase emphasizes higher-quality multimodal supervision and longer-context capability refinement.

The technical consequence of this recipe is that the model learns more than a raw association between audio and text. It learns an operational interface between them. That interface is later reused in three directions: ASR maps audio evidence into text tokens, TTS maps text-side semantics into audio tokens, and Realtime couples listening, reasoning, and response generation under turn-level latency constraints. In this sense, pretraining is not merely background context for the rest of the report; it is the central mechanism that explains why all three specializations can share one backbone.




\section{ASR Specialization}
\label{sec:asr}

\begin{figure}[tbp]
\centering
\includegraphics[width=\linewidth]{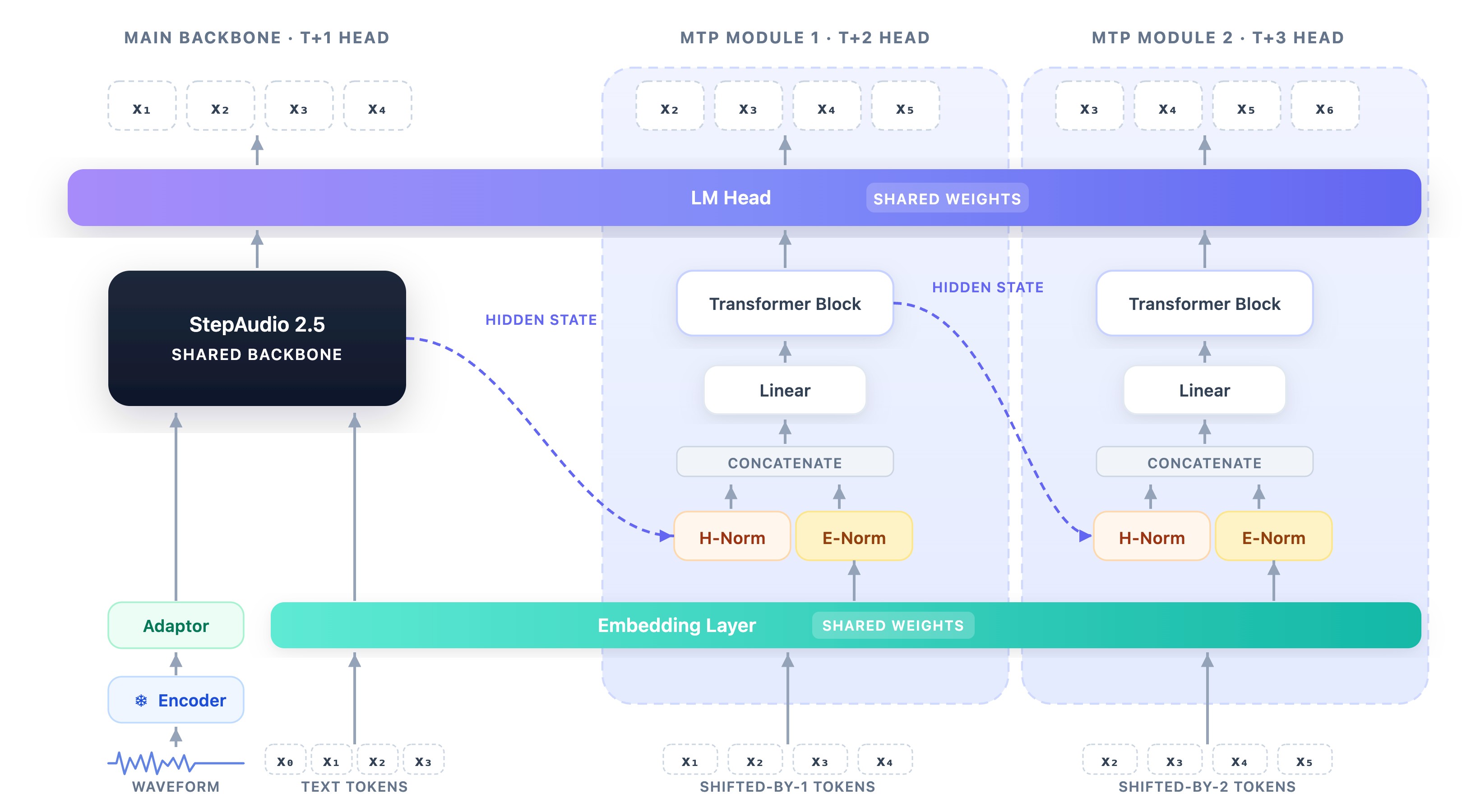}
\caption{ASR architecture in StepAudio 2.5. The shared encoder-adaptor-decoder backbone is augmented with parallel future-token branches, making decoding substantially more efficient while preserving autoregressive verification.}
\label{fig:asr_architecture}
\end{figure}

StepAudio 2.5 ASR follows the StepAudio encoder-adapter-decoder pattern, augmented with an \mtpmodel\ head that proposes verifiable future transcript tokens, as shown in Figure~\ref{fig:asr_architecture}. 
At decoding position $t$, the main branch predicts the next transcript token $x_{t+1}$. The $h$-th MTP branch predicts $x_{t+1+h}$ for $h\in\{1,\ldots,5\}$, so one forward step produces a six-token proposal. During inference, the proposal is accepted only as a verified prefix: once a future token disagrees with the normal decoding path, later proposed tokens are rejected and decoding continues autoregressively from the accepted prefix. This verification mechanism ensures that MTP acts strictly as an acceleration primitive. 

Each MTP block receives the hidden state from the previous branch and a shifted token embedding. The two inputs are normalized, concatenated, projected back to the decoder hidden size, and processed by a decoder-style Transformer block. All branches share the same embedding layer and vocabulary output head as the main decoder. 

\subsection{Training Pipeline}

\textbf{ASR SFT} Supervised fine-tuning first turns the model into a reliable autoregressive recognizer using both short-form and long-form data. Training examples are packed into a 32K-token sequence budget. SpecAugment-style time and frequency masking~\citep{park2019specaugment} is applied to the acoustic features. Throughout this stage, the audio encoder remains frozen, while the adapter and language decoder are optimized for 10K steps with a peak learning rate of $2\times10^{-5}$, a global batch size of 32, 100 warmup steps, and cosine decay to $1\times10^{-6}$.

\textbf{MTP Training} After the base recognizer has well converged, MTP is introduced as a lookahead proposal module through a staged optimization recipe: frozen-branch alignment and joint calibration. 

\begin{itemize}
    \item \textbf{Frozen-branch alignment.} Five MTP blocks are appended to the converged ASR decoder. The Transformer layer within each block is initialized from the last decoder layer to inherit a strong linguistic prior, while the branch-specific projections are newly initialized. In this stage, only the MTP blocks are optimized with a peak learning rate of $2\times10^{-4}$, while all other modules including the shared token embeddings and LM head remain frozen. 
    \item \textbf{Joint calibration.} Once the branches have aligned with the ASR distribution, the adapter and LLM decoder are unfrozen for joint optimization with a lower learning rate of $2\times10^{-5}$. This stage reduces the residual mismatch between the backbone states and the lookahead branches, turning MTP into a calibrated proposal mechanism. 
\end{itemize}

Both stages inherit the 32K sequence budget, 32 global batch size, and 10K-step training horizon.
During training, the main branch predicts the next token $x_{t+1}$ at position $t$, while the $h$-th MTP branch targets the future token $x_{t+1+h}$ for $h\in\{1,\ldots,H\}$. The branch weights are exponentially decayed to reflect the serial dependency of MTP:
\[
w_h = \frac{\alpha^{h-1}}{\sum_{j=1}^{H}\alpha^{j-1}}, \quad H=5, \quad \alpha=0.9.
\]
At each position $t$, the final objective combines the standard next-token loss with the weighted MTP losses:
\[
\mathcal{L}_t = \mathrm{CE}(p_t, x_{t+1}) + \sum_{h=1}^{H} w_h \mathrm{CE}(p_{t,h}, x_{t+1+h}),
\]
where $p_t$ and $p_{t,h}$ are the distributions from the main and auxiliary branches, respectively.

\subsection{Data}
\textbf{Short-form supervised data.} The short-form SFT set comprises approximately 100K hours of audio, integrating major public corpora with inhouse datasets. The mixture covers a wide spectrum of linguistic and acoustic variations, including Mandarin, English, and frequent code-switching utterances. To handle real-world complexity, the data also spans various vertical domains rich in professional terminologies, as well as challenging acoustic environments such as far-field recording and high-noise scenarios. Each sample in this set has a maximum duration of 30 seconds.



\begin{figure}[tbp]
\centering
\includegraphics[width=\linewidth]{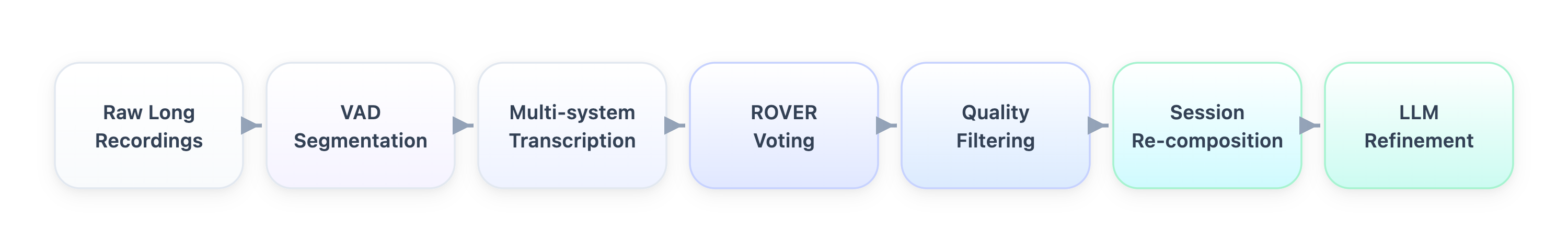}
\caption{Long-form ASR data construction pipeline. The process transitions from individual clip transcription to global session-level refinement to ensure both accuracy and consistency.}
\label{fig:longdata}
\end{figure}

\textbf{Long-form pseudo-labeled data.} While short-form data ensures utterance-level precision, long-duration recordings are essential for teaching the model to maintain contextual consistency. To support this capability, the training recipe curates a 50K-hour long-form dataset using a multi-system verification pipeline designed to provide reliable session-level supervision, as shown in Figure~\ref{fig:longdata}. Raw recordings are first segmented by Voice Activity Detection (VAD) into speech clips with a maximum duration of 30 seconds. Each clip is transcribed independently by three ASR systems to obtain multiple candidate hypotheses. To focus the subsequent fusion on genuine recognition errors, these hypotheses undergo surface-form normalization to unify formatting, casing, and punctuation. The normalized streams are then aligned and fused via Recognizer Output Voting Error Reduction (ROVER)~\citep{fiscus1997rover}, with voting performed at token level. Tokens are accepted only when supported by at least two systems, while non-consensus positions are marked as disagreements. The segment-level disagreement rate $\hat{e}$ serves as a proxy for label reliability:
\[
\hat{e}=\frac{\#\mathrm{disagreed\ positions}}{\#\mathrm{text\ units}}
\]
Clips with $\hat{e}>0.05$ are discarded to maintain high training signal fidelity. Passing neighbor segments are then concatenated into long-form training samples. Finally, an LLM-based refinement stage restores punctuation, performs inverse text normalization, and ensures cross-segment consistency by harmonizing recurring terminology and entities across the full session.


\subsection{Evaluation}

The evaluation of StepAudio 2.5 ASR focuses on three primary objectives: recognition accuracy across diverse languages, native long-form transcription capability, and inference efficiency under production-scale serving. We compare the system against several competitive baselines, including VibeVoice-ASR~\citep{peng2026vibevoiceasrtechnicalreport}, FunASR-Nano~\citep{an2025funasrtechnicalreport}, Doubao-ASR-2603~\citep{bai2024seed} and Qwen3-ASR-1.7B~\citep{shi2026qwen3asrtechnicalreport}. To ensure a fair comparison, all models are deployed in a local environment using a single NVIDIA H800 GPU with single-concurrency serving, except for Doubao-ASR-2603, which is only accessible through the official API. For baseline models that do not natively support long-form audio like FunASR-Nano, VAD is used to segment recordings into clips with a maximum duration of 30 seconds. Recognition benchmarks draw on AISHELL-1~\citep{bu2017aishell}, AISHELL-2 (iOS test)~\citep{du2018aishell2}, WenetSpeech~\citep{zhang2022wenetspeech}, FLEURS~\citep{conneau2022fleurs}, LibriSpeech~\citep{panayotov2015librispeech}, Common Voice~\citep{ardila2020commonvoice}, VoxPopuli cleaned AA~\citep{artificialanalysis2026voxpopulicleaned}, and Earnings22 cleaned AA~\citep{artificialanalysis2026earnings22cleaned}. Long-form evaluation includes LibriSpeech long variants, Earnings22 cleaned AA, and WenetSpeech testnet long~\footnote{
WenetSpeech testnet long is constructed by merging adjacent WenetSpeech testnet segments into extended recordings. We release \url{https://github.com/lawlict/wenetspeech-testnet-long.git} for corpus generation.}.


\textbf{Recognition performance.} Table~\ref{tab:asr_results} shows that StepAudio 2.5 ASR establishes a new performance frontier for LLM-based ASR. On Chinese benchmarks, the model improves the average CER to 2.97\%, with a notable reduction in error rate on AISHELL-1 to 0.71\% and a competitive 2.63\% on FLEURS zh. On English, the model reduces the average WER to 3.68\%, outperforming competitive baselines and showing particular strength on LibriSpeech clean (1.38\%) and VoxPopuli cleaned AA (2.76\%). Long-form transcription is where decoder context and linguistic depth matter simultaneously. The model reaches the best average long-form error rate of 3.70\%, representing a significant improvement over Qwen3-ASR-1.7B. The model is especially effective on the long LibriSpeech variants, where the native 32K context window allows it to maintain consistent recognition without the boundary errors typical of segmentation-based approaches.

We also compare StepAudio 2.5 ASR against the base ASR model after SFT but before MTP training. Across Chinese, English, and long-form benchmarks, the addition of MTP-5 leaves recognition accuracy essentially unchanged, with average fluctuations within 0.06 absolute points. This stability is the result of the staged training recipe and the autoregressive verification process, which ensures that the final transcript is always determined by the verified path.


\begin{table}[tbp]
\centering
\small
\setlength{\tabcolsep}{3.2pt}
\caption{ASR results on Chinese, English, and long-form benchmarks (Error Rate, \%). Lower is better. The second-best results are underlined.}
\label{tab:asr_results}
\begin{adjustbox}{max width=\textwidth}
\begin{tabular}{llcccccc>{\color{gray}\arraybackslash}c}
\toprule
Category & Test set & VibeVoice-ASR & FunASR-Nano & Doubao-ASR-2603 & Qwen3-ASR-1.7B & StepAudio 2.5 ASR & \textcolor{gray}{\makecell{StepAudio 2.5 ASR\\w/o MTP training}} \\
\midrule
\multirow{6}{*}{Chinese}
& AISHELL-1 & 5.19 & 1.88 & 2.07 & \underline{1.49} & \textbf{0.71} & \textcolor{gray}{0.79} \\
& AISHELL-2 ios & 5.10 & 2.61 & 2.70 & \underline{2.50} & \textbf{2.29} & \textcolor{gray}{2.30} \\
& WenetSpeech testnet & 14.79 & 5.30 & \textbf{4.03} & \underline{4.44} & 4.54 & \textcolor{gray}{4.57} \\
& WenetSpeech testmeeting & 17.09 & 5.31 & 5.09 & \textbf{4.66} & \underline{4.70} & \textcolor{gray}{4.73} \\
& FLEURS zh & 8.77 & 3.19 & 2.83 & \underline{2.74} & \textbf{2.63} & \textcolor{gray}{2.63} \\
\cmidrule(lr){2-8}
& Average & 10.19 & 3.66 & 3.34 & \underline{3.17} & \textbf{2.97} & \textcolor{gray}{3.00} \\
\midrule
\multirow{6}{*}{English}
& LibriSpeech clean & 2.30 & 1.80 & 2.94 & \underline{1.69} & \textbf{1.38} & \textcolor{gray}{1.40} \\
& LibriSpeech other & 5.79 & 4.43 & 5.98 & \underline{3.57} & 3.16 & \textcolor{gray}{3.14} \\
& Common Voice v11 en & 20.03 & 11.05 & 14.06 & \textbf{7.50} & \underline{7.57} & \textcolor{gray}{7.62} \\
& FLEURS en & 5.20 & 4.96 & 6.74 & \textbf{3.23} & \underline{3.55} & \textcolor{gray}{3.74} \\
& VoxPopuli cleaned AA & \textbf{2.38} & 3.97 & 3.61 & 3.28 & \underline{2.76} & \textcolor{gray}{3.23} \\
\cmidrule(lr){2-8}
& Average & 7.14 & 5.24 & 6.67 & \underline{3.85} & \textbf{3.68} & \textcolor{gray}{3.83} \\
\midrule
\multirow{5}{*}{Long-form}
& LibriSpeech clean long & \underline{1.66} & 2.34 & 2.81 & 1.95 & \textbf{1.27} & \textcolor{gray}{1.27} \\
& LibriSpeech other long & 3.48 & 4.89 & 5.59 & \underline{3.81} & 2.90 & \textcolor{gray}{2.81} \\
& WenetSpeech testnet long & 8.73 & 4.74 & \textbf{3.72} & 4.15 & \underline{4.09} & \textcolor{gray}{4.09} \\
& Earnings22 cleaned AA & \textbf{5.62} & 10.38 & 12.33 & 6.90 & \underline{6.52} & \textcolor{gray}{6.34} \\
\cmidrule(lr){2-8}
& Average & 4.87 & 5.59 & 6.11 & 4.20 & \underline{3.70} & \textcolor{gray}{3.63} \\
\bottomrule
\end{tabular}
\end{adjustbox}
\end{table}

\textbf{Decoding efficiency.} Table~\ref{tab:rtf} evaluates the deployment objective directly. The Real-Time Factor is measured on 100 clips of 30 seconds each. StepAudio 2.5 ASR reaches an exceptionally low RTF of 0.0053, faster than the Qwen3-ASR-1.7B baseline despite using a larger decoder. It is also substantially faster than VibeVoice-ASR, FunASR-Nano, and Doubao-ASR-2603 under the same serving setup. This is the key systems consequence of MTP for ASR: decoder scale no longer translates linearly into token-by-token latency, because most steps emit several verified transcript tokens.

\begin{table}[htb]
\centering
\small
\caption{RTF comparison.}
\label{tab:rtf}
\begin{tabular}{lccccc}
\toprule
Model & VibeVoice-ASR & FunASR-Nano & Doubao-ASR-2603 & Qwen3-ASR-1.7B & StepAudio 2.5 ASR \\
\midrule
RTF & 0.1039 & 0.0591 & 0.0640 & 0.0094 & \textbf{0.0053} \\
\bottomrule
\end{tabular}
\end{table}

\textbf{MTP acceptance behavior.} Experiments evaluate the strict per-position acceptance rate on the WenetSpeech meeting set~\citep{zhang2022wenetspeech}. To determine the optimal lookahead horizon, three configurations are compared: MTP-3, MTP-5, and MTP-7. Results in Table~\ref{tab:acceptance} reveal two primary trends. First, the acceptance rates of earlier positions are nearly invariant to the total number of branches, indicating that each MTP head learns a stable, independent prediction task. Second, starting from the second position, the acceptance rate decays at a consistent factor of approximately $0.9$ per branch. While increasing branches from three to five yields a substantial 39\% gain in average accepted length, the additional step to MTP-7 provides a more modest improvement about 22\%. This diminishing return is driven by the high failure rates of the sixth and seventh positions, which frequently trigger KV cache rollbacks and interrupt the decoding stream, and finally offset the marginal utility of a longer lookahead. Consequently, MTP-5 represents a deliberate choice for the optimal efficiency-complexity trade-off.

\begin{table}[htb]
\centering
\small
\setlength{\tabcolsep}{8pt}
\caption{Strict per-position MTP acceptance rate and average accepted length.}
\label{tab:acceptance}
\begin{adjustbox}{max width=\textwidth}
\begin{tabular}{lccccccc|c}
\toprule
Config & 1st & 2nd & 3rd & 4th & 5th & 6th & 7th & Avg. Length \\
\midrule
MTP-3 & 0.96 & 0.88 & 0.80 & -- & -- & -- & -- & 3.6 / 4 \\
MTP-5 & 0.95 & 0.88 & 0.80 & 0.71 & 0.64 & -- & -- & 5.0 / 6 \\
MTP-7 & 0.96 & 0.88 & 0.80 & 0.72 & 0.65 & 0.59 & 0.53 & 6.1 / 8 \\
\bottomrule
\end{tabular}
\end{adjustbox}
\end{table}

\textbf{Insight.} The ASR branch suggests a useful general lesson for multimodal systems: grounded generation tasks can sometimes be accelerated more aggressively than free-form text generation, precisely because the external modality reduces semantic branching. In other words, grounding is not only a source of information; it is also a source of algorithmic structure.

\section{TTS}
\label{sec:tts}
Compared with our ASR and Realtime speech models, the distinctive aspect of StepAudio 2.5 TTS is that it completely eliminates the encoder-adapter module and relies solely on the LLM backbone for modeling. Audio tokens are treated as a new “language” within the language modeling framework, allowing speech synthesis to be reformulated as a pure next-token prediction (NTP) task. Under this paradigm, the key challenge becomes to learn effective alignment between text and audio representation spaces.
\subsection{Training Pipeline}
\label{sec:pipeline}

After pre-training, the model establishes a shared representation space between text and audio, enabling unified cross-modal modeling. Building upon this foundation, SFT further aligns text descriptions with their corresponding audio token sequences, allowing the model to generate speech conditioned on natural language instructions. Reinforcement learning is then introduced to improve alignment in scenarios where instructions become more complex, abstract, or highly context-dependent, thus improving the model’s ability to faithfully capture nuanced semantic intent in generated speech.

\textbf{SFT}
To enable the model to support both global- and inline-level controllability over arbitrary speakers, we adopt zero-shot voice cloning TTS as the primary objective during supervised fine-tuning. In the first stage, we conduct large-scale zero-shot TTS training with global instruction supervision, enabling the model to learn coarse-grained control over speaker characteristics, speaking style, and overall prosodic attributes. Building upon this capability, we further train the model using high-quality in-house speech data annotated with both global and inline instructions, thereby enabling fine-grained control at both the utterance and segment levels. Through this two-stage SFT pipeline, the model achieves an initial alignment between textual control instructions and their corresponding audio realizations, supporting both global instruction following and inline expressive control in TTS generation.

\textbf{Reinforcement Learning}
During the RL stage, we apply human feedback reinforcement learning (RLHF) to further align the generated audio tokens with textual descriptions based on human preferences. This process improves the model’s ability to interpret complex and expressive instructions, while simultaneously improving the naturalness, expressiveness, and overall perceptual quality of the synthesized speech. 

We first train a Generative Reward Model (GRM), denoted as $r_\phi$. For each prompt $x$, the training data provides a high-quality reference response $y^*$. During training, the policy model $\pi_\theta$ generates a candidate response $y$, and the GRM evaluates the relative quality of $y$ against the reference response $y^*$ under the same prompt $x$, producing a pairwise scalar preference score. The final reward used for policy optimization is obtained by applying a reward-shaping transformation to this score:

\begin{equation}
r_{hf}(x, y, y^*) = s\!\left(r_\phi(x, y, y^*)\right),
\end{equation}

where $r_\phi(x, y, y^*)$ denotes the GRM score relative to the reference response and $s(\cdot)$ denotes the transformation of the reward.

\subsection{SFT Data}
\label{sec:SFT-data}
Our SFT data consist of two categories: model-synthesized data and recorded speech data. All data used for global instruction control is synthesized using the Step-Audio-EditX~\citep{yan2025stepaudioeditxtechnicalreport} model. Owing to the strong compositional editing capabilities of Step-Audio-EditX across diverse speaking styles and emotional attributes, we are able to generate large-scale speech data with rich and highly diverse stylistic and emotional variations for global instruction supervision.

In contrast, the recorded speech data is primarily used for joint global and inline control, enabling fine-grained hierarchical expressive modeling. The detailed data construction pipeline is described below.

For internal recordings with clear dialogue context, speaker information, or available scripts, we adopt the annotation pipeline of Emotional-Context-Speech \footnote{https://huggingface.co/Insects/Emotional-Context-Speech}. The main change is the annotation target. Instead of producing structured context-aware labels, we generate two forms of natural-language supervision for zero-shot TTS: (1) a global control description for the whole utterance, and (2) inline expression descriptions for local text spans.

Following the source pipeline, we first transcribe the recordings with Whisper-Large-v3. We then use the Montreal Forced Aligner to obtain word-level timestamps and segment the recordings into utterance-level samples. We remove samples with severe alignment errors, incomplete transcripts, or very short duration. For each retained utterance, we collect contextual metadata, including dialogue history or script context. This contextual information is used during annotation.

Following Emotional-Context-Speech, we extract and discretize prosodic features, including F0, speech rate, pause statistics, spectral centroid, root mean square (RMS) energy, MFCC variance, and harmonic-to-noise ratio (HNR). We concatenate these tokenized acoustic features with the transcript and contextual metadata, and feed them to the annotating LLM.

Given the transcript, contextual metadata, and quantized acoustic features, the LLM produces two annotation outputs. The first is a global control description, which summarizes the overall speaking style, prosodic pattern, and affective state of the whole utterance. The second is an inline expression description, which inserts local directives into the text to mark span-level expressive behavior. These two outputs provide supervision for utterance-level control and mixed expression control in zero-shot TTS.

Before model training, we further verify the transcription accuracy of the segmented utterances by cross-checking their transcripts with the outputs of our ASR models. Segments with substantial transcription mismatch are discarded.

\subsection{Evaluation}
Effectively evaluating the diverse capabilities of modern TTS systems remains a challenging problem. Traditional objective metrics, such as character error rate (CER) and speaker similarity, often exhibit inherent bias when applied to LLM-based speech generation models. For instance, ASR-based metrics tend to become unreliable in the presence of rich paralinguistic phenomena, while embedding-based speaker verification models typically discard high-frequency acoustic details and fail to accurately capture similarities in prosody, speaking style, and expressive characteristics.

Similarly, LLM-as-a-judge approaches still struggle to reliably assess prosodic quality and complex emotional expression. Subjective MOS evaluation also presents significant limitations, as it requires highly trained annotators and often suffers from inconsistencies in scoring criteria across evaluators.

Considering these limitations, we adopt an arena-style pairwise evaluation framework, in which models are compared via pairwise preference judgments, and their overall performance is measured by aggregated win rates. To ensure evaluation reliability, we invest substantial effort in standardizing the evaluation protocol and improving inter-rater consistency among human evaluators.

Specifically, we proceed as follows: (1) We first conduct a listening sensitivity screening using a small set of audio samples to select qualified evaluators. Once the evaluation task begins, the set of evaluators remains fixed, and all evaluations must be completed continuously within the same evaluation period. (2) During the evaluation process, we ensure randomness in both the selection of model audio pairs and the ordering of evaluation positions, and we additionally require evaluators to provide reasons for their preference judgments. (3) We perform periodic spot checks during the evaluation process and intervene promptly when significant deviations are observed to maintain inter-rater consistency. After the full evaluation is completed, we further review cases with large discrepancies across evaluators and conduct additional verification to ensure the reliability of the final results.

\begin{figure}[h]                                    
\centering                               
\includegraphics[width=1.0\linewidth]{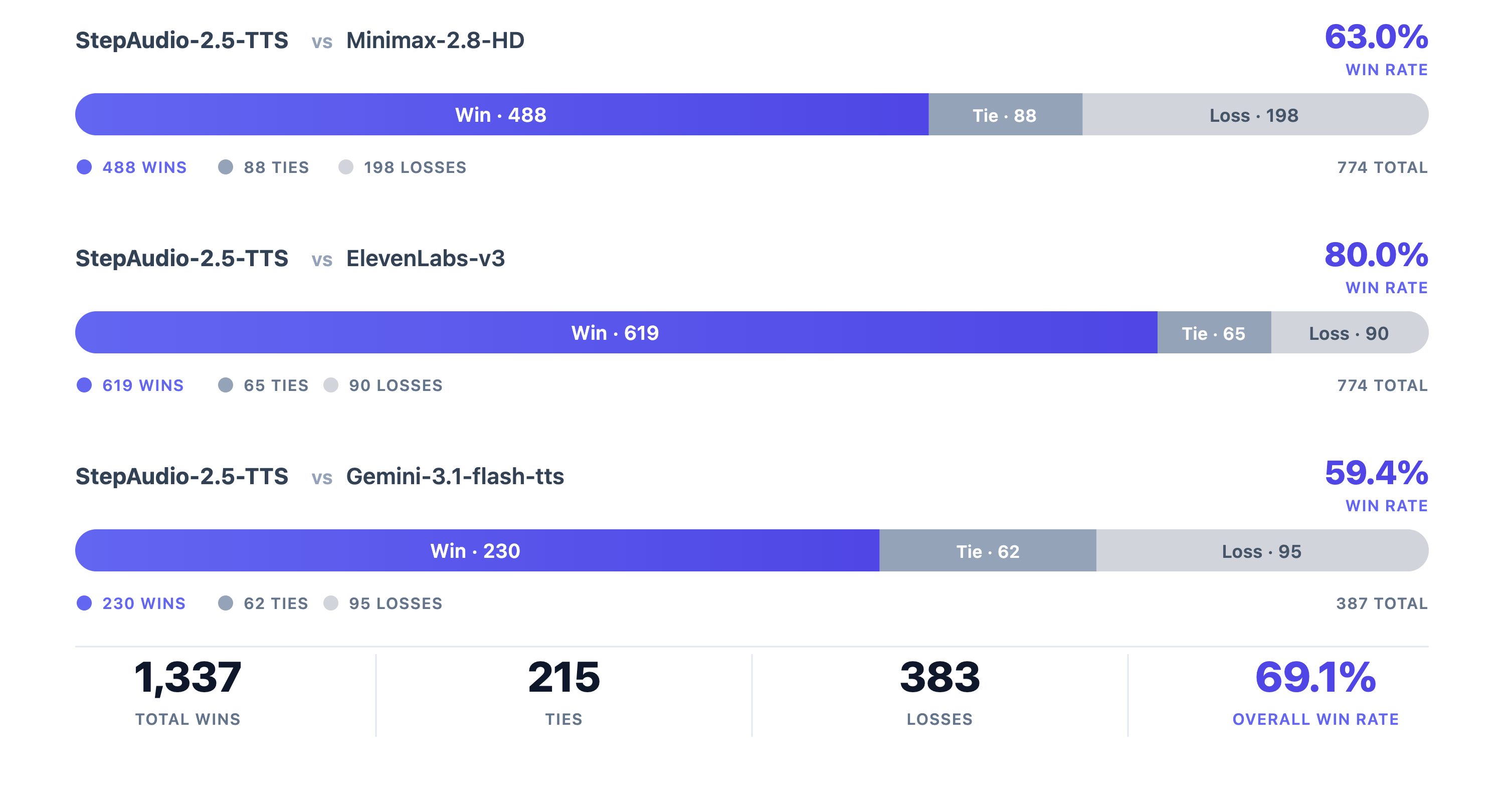}   
\caption{Arena Win Rates of \textit{StepAudio-2.5-TTS}.} 
\label{fig:tts_arena}                                
\end{figure}  

Finally, we select three leading models with controllable generation capabilities—\textit{MiniMax-2.8-HD}, \textit{Elevenlabs-v3}, and \textit{Gemini-3.1-Flash-TTS}. For each model, we adopt its officially recommended optimal voice preset and conduct arena-based evaluation using 774 prompts.

The results in Fig.~\ref{fig:tts_arena} show that \textit{StepAudio-2.5-TTS} achieves 67.6\% overall win rate in pairwise evaluations against three strong TTS baselines, with consistent gains across all comparisons.

\section{Realtime Specialization}
\label{sec:realtime}

The StepAudio 2.5 foundation is specialized into three branches: an ASR branch, a TTS branch, and the realtime spoken interaction branch detailed in this section. \steprealtime\ inherits the core foundation architecture without modification---an audio encoder, an audio adaptor that projects acoustic representations into the decoder's hidden space, and a large decoder that produces an explicit latent reasoning trace before generating a response.

Targeting multi-turn spoken interaction under stringent latency constraints introduces distinct challenges compared to standard speech-to-text tasks:
\begin{itemize}
    \item \textbf{Conversational Coherence:} Maintaining topical context, stylistic coherence, and dialogue state across extended interactions.
    \item \textbf{Persona Consistency:} Adhering to specific personality traits and delivery styles under diverse and adversarial user inputs.
    \item \textbf{Paralinguistic Sensitivity:} Understanding and appropriately responding to non-verbal cues such as hesitation, laughter, sighs, and pacing changes.
    \item \textbf{Reward Sparsity:} Unlike discrete QA tasks, conversational attributes (e.g., naturalness, emotional fit) lack a single ground-truth target, making them difficult to optimize solely via verifiable reward signals.
\end{itemize}
To address these challenges, we introduce a specialized training pipeline relying on data composition and a staged optimization recipe, without structural changes to the architecture.

\subsection{Training Pipeline}
The training pipeline systematically instills conversational capabilities while preserving the foundation model's perceptual and reasoning abilities. It consists of three main stages: audio-centric mid-training, multi-stage Supervised Fine-Tuning (SFT), and Reinforcement Learning from Human Feedback (RLHF).

\subsubsection{Audio-Centric Mid-Training}
Inherited directly from the foundation model, this phase equips the model with robust audio-grounded perception and long-form reasoning capabilities. It serves as the baseline before dialogue-specific behaviors are introduced.

\subsubsection{Progressive Supervised Fine-Tuning (SFT)}
\label{sec:realtime_sft}
The SFT phase serves as the primary vehicle for transforming the base model into a natural conversationalist. Rather than treating this as a monolithic training step, we adopt a progressive curriculum that systematically injects interactive capabilities across three core dimensions:

\begin{itemize}
    \item \textbf{Conversational Alignment:} The model is first calibrated for multi-turn interactions. Using instruction-rich dialogue data, we train it to maintain turn-level continuity, handle spoken-language artifacts (e.g., disfluencies, mid-utterance interruptions), and favor colloquial, prosody-friendly responses over formal text.
    \item \textbf{Persona and Stylistic Control:} To support diverse interaction archetypes, we introduce scalable, persona-conditioned data. By expanding a curated seed into a massive feature matrix of personality traits and verbal habits, we train the policy to condition jointly on persona specifications and dialogue history. This enables compositional generalization, allowing the model to adapt its delivery descriptors and non-verbal vocalizations (e.g., laughter, sighs) to unseen persona combinations.
    \item \textbf{Paralinguistic Sensitivity:} Using real spoken interactions, the model is trained to recognize subtle paralinguistic cues. It learns to register these cues in its latent reasoning trace and dynamically adjust its response tone and pacing. Consequently, the model synthesizes the static external persona (\emph{who is speaking}) with real-time paralinguistic reading (\emph{how the user is speaking}).
\end{itemize}

To prevent catastrophic forgetting and stylistic drift as these capabilities are introduced, we employ a \textbf{dynamic rehearsal schedule}. Throughout the SFT phase, interaction-specific data is continuously interleaved with general-purpose instruction data and reasoning tasks based on validation metrics, ensuring the model retains its foundational reasoning while mastering complex dialogue behaviors.

\subsubsection{RLHF with Generated Rewards}
To bridge the residual gap in dialogue quality where no single SFT demonstration is optimal, we apply RLHF using a PPO-style objective~\citep{schulman2017proximal} with KL regularization. A generative reward model scores candidates against reference responses, utilizing explicit interaction rubrics to guide instruction-sensitive aspects, while standard preference comparisons govern overall conversational naturalness.

Training is conducted on a mixture of multi-turn dialogues and single-turn prompts: multi-turn data encourages the policy to maintain consistency across exchanges, while single-turn prompts provide capacity for longer-form reasoning and richer preference articulation. Preference comparisons serve as the primary reward signal, capturing overall response quality. Rubric scores complement this signal on instruction-sensitive aspects, particularly those where consistency matters most, including maintaining coherence across turns and remaining faithful to earlier user content. Unlike conventional scalar reward models, the generative reward model captures finer-grained aspects of human preference, providing a richer training signal for policy alignment.

\subsection{Data}
The SFT data for \steprealtime\ is organized into three complementary streams that mirror the staged objective in Section~\ref{sec:realtime_sft}. The conversational backbone consists of multi-turn dialogues drawn from natural spoken interaction, filtered to favor turn-to-turn continuity, elliptical or disfluent phrasing, and mid-utterance revisions, with written-style responses down-weighted so that the policy is anchored to a spoken register.

Layered on top is persona-conditioned dialogue at scale. Starting from more than 10{,}000 native personas authored end-to-end and validated by human reviewers, an algorithmic fission procedure recombines orthogonal attributes—personality, verbal habits, emotional boundaries, and interaction archetypes—into a million-scale persona matrix, and each synthesized persona is paired with dialogues drawn from a million-scale real-scenario corpus so that persona attributes are grounded in realistic interaction contexts.

A third stream exposes the model to paralinguistic cues. These dialogues carry an atmosphere descriptor that governs speaking rate, stress, and subtext, alongside cue labels covering hesitation, light laughter, sigh, breath, change of pace, and falling intonation. Interleaved throughout is a general-capability mixture inherited from mid-training that preserves reasoning ability, and the full corpus passes through a unified pipeline that checks in-character consistency, cross-validates annotations, and removes near-duplicates introduced by fission.

\subsection{Evaluation}

Because realtime interaction quality depends on properties that transcript-level metrics do not capture, we evaluate \steprealtime\ in a fully interactive setting that combines subjective human evaluation conducted through mobile-app sessions with objective API-based evaluation across general dialogue, in-car dialogue, dialogue understanding, and audio-question answering. The five suites are:
\begin{itemize}
    \item \textbf{Step-Dialogue-Human-Eval}: Subjective mobile-app evaluation for general dialogue scenarios.
    \item \textbf{step\_Dialogue\_general}: Objective API evaluation for general dialogue.
    \item \textbf{step-Dialogue-car}: Objective API evaluation for in-car dialogue scenarios.
    \item \textbf{Step-Dialogue-Understanding}: 87 diverse audio samples testing the model's ability to infer speaker acoustic features (e.g., age, gender, speech rate) directly from the audio signal.
    \item \textbf{Step-SPQA}: An 11-category audio-question/audio-answer benchmark introduced in Step-Audio 2.
\end{itemize}


\begin{figure}[h]                                    
\centering                               
\includegraphics[width=1.0\linewidth]{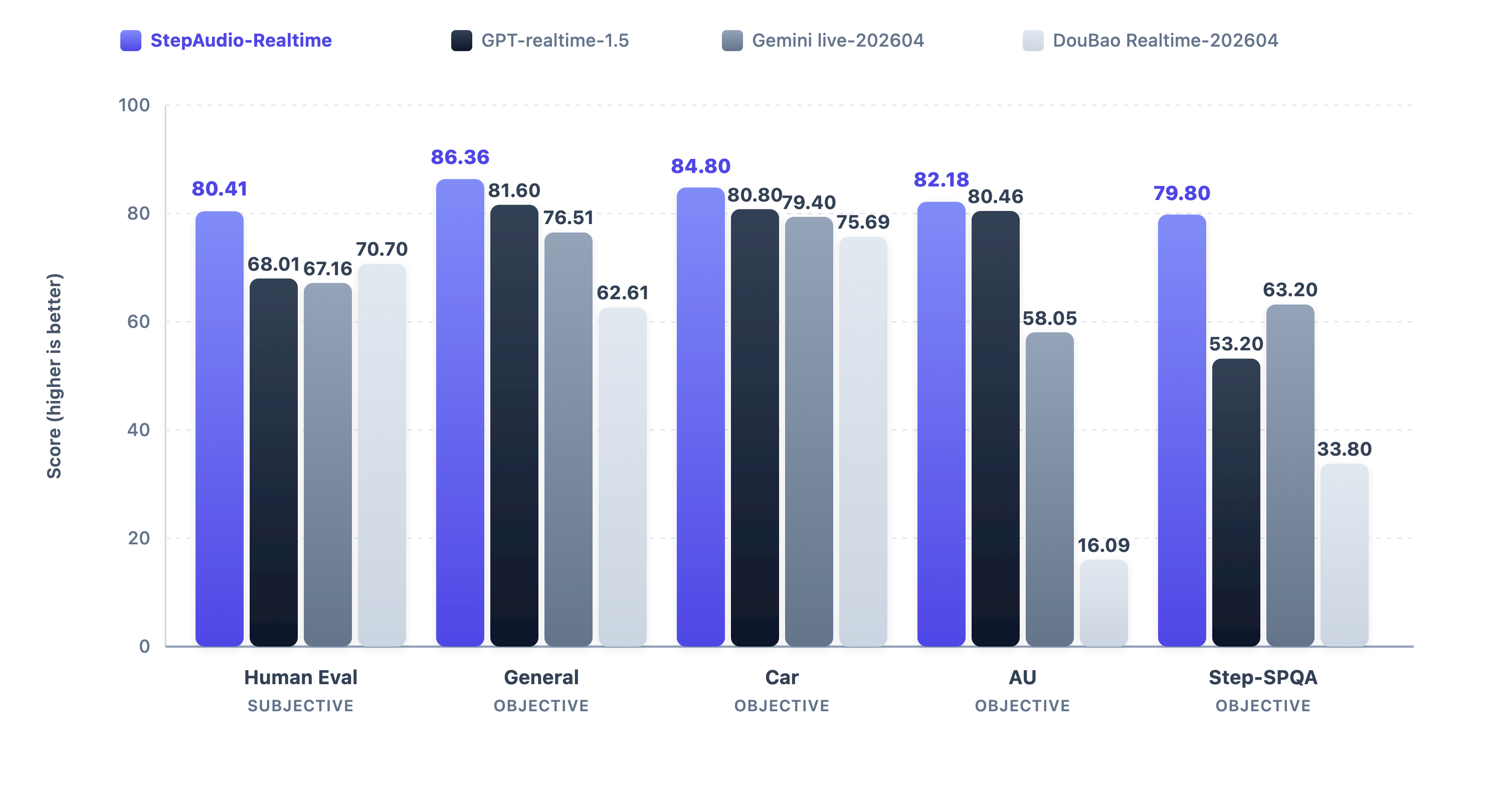}   
\caption{Realtime interaction evaluation. Higher is better. Best results are in bold.} 
\label{fig:realtime-eval}                                
\end{figure}  

\textbf{Results Analysis:}
As shown in Figure~\ref{fig:realtime-eval}, \steprealtime\ consistently outperforms competitive baselines across all five suites.
Notably, it achieves a +10.0 margin on the subjective human evaluation compared to the next-best system, validating the efficacy of our persona and naturalness conditioning. Furthermore, the +16.6 margin on Step-SPQA and strong performance on Step-Dialogue-Understanding indicate that the paralinguistic conditioning enhances acoustic comprehension without degrading general reasoning. The concurrent improvements in both subjective conversational quality and objective audio understanding demonstrate that our rehearsal schedule effectively balances specialized interaction training with foundational capabilities.

\section{Conclusion}
\label{sec:conclusion}

This report presents StepAudio 2.5 as a unified audio-language foundation with three downstream specializations. The shared backbone is learned through a staged multimodal curriculum that aligns speech and text, extends the token interface to audio, scales unified multimodal training, and refines the model with high-quality long-context data. On top of that backbone, the ASR branch turns speech recognition into a particularly favorable application of verifiable multi-token decoding, the TTS branch turns speech generation into a problem of semantic-to-audio alignment strengthened by context-rich supervision and reinforcement learning, and the Realtime branch extends the same foundation to low-latency spoken dialogue with persona stability and paralinguistic sensitivity. Taken together, these capabilities indicate that StepAudio 2.5 is best understood not as a collection of isolated speech endpoints, but as a shared foundation whose recognition, synthesis, and realtime interaction abilities emerge through different optimization and deployment regimes.

\section*{Authors}
The contributors are listed in alphabetical order.

\vspace{0.5em}
\textbf{Core Contributors:}
Bin Lin,
Bo Zhao,
Boyong Wu,
Chao Yan,
Chen Wu,
Cheng Yi,
Chengyuan Yao,   
Daijiao Liu,
Fei Tian,
Feng Tian,   
Haiyang Sun,
Haoyang Zhang,
Jiangjie Zhen,   
Jinglan Gong,
Jun Chen,
Li Xie,   
Peilin Li,   
Peng Yang,
Pengfei Tan,
Qingjian Lin,
Runze Li,
Shenghua Hu,   
Siyi Zhou,
Wenwen Qu,   
Xiangyu Li,   
Xiangyu Tony Zhang,
Xuerui Yang,   
Yang Yang,   
Yechang Huang,
Yu Fu,
Yuchu Luo,   
Yuxin Li,
Yuxin Zhang,
Zhengyan Sheng

\vspace{0.5em}
\textbf{Contributors:}
Brian Li,
Chang Zeng,
Changlin Zhang,
Chen Geng,
Chenghao Dong,
Chengli Feng,
Dan Zhou,
Danni Wan,
Di Chen,
Die Zhang,
Dongqing Pang,
Guanglong Yang,
Guoqiang Hu,
Huangxi Zhu,
Jianzheng Gao,
Jinghua Liang,
Jinmei Wan,
Junjie Yuan,
Kang An,
Lei Lei,
Limin Zhong,
Lun Cai,
Mengqiang Ren,
Min Xu,
Mingliang Li,
Mingxiao Li,
Na Wang,
Qiang Tong,
Qiaoling Huang,
Qingfu Du,
Rui Wang,
Shengchen Zhou,
Shi Qiu,
Shihao Peng,
Shiliang Yang,
Siqi Tu,
Tianjiao Deng,
Ting Xu,
Tong Wang,
WeiMing Niu,
Wuxun Xie,
Xianwei Zhang,
Xianyu Feng,
Xiaojia Liu,
Xing Chen,
Xiongbin Wu,
Yan Wu,
Yang Li,
Yi Liu,
Yifan Zhang,
Yile Liu,
Yongshen Long,
Yu Luo,
Yuanhao Ding,
Yuhao Wang,
Yuhe Yin,
Yunfang Xu,
Yuxiang Yang,
Zhiguo Huang,
Zhiyue Wu,
Zichao Li,
Zichao Zhou

\vspace{0.5em}
\textbf{Sponsers:}
Daxin Jiang,
Future Li,
Gang Yu,
Xiangyu Zhang,
Yibo Zhu

\setlength{\bibsep}{0.5\baselineskip}

\end{document}